# Comment on Photon Exchange Interactions

J.D. Franson
Johns Hopkins University, Applied Physics Laboratory, Laurel, MD 20723

*Abstract:*

We previously suggested that photon exchange interactions could be used to produce nonlinear effects at the two-photon level, and similar effects have been experimentally observed by Resch et al. (quant-ph/0306198). Here we note that photon exchange interactions are not useful for quantum information processing because they require the presence of substantial photon loss. This dependence on loss is somewhat analogous to the postselection required in the linear optics approach to quantum computing suggested by Knill, Laflamme, and Milburn [Nature **409**, 46 (2001)].

Some time ago, we suggested that photon exchange interactions could be used to produce nonlinear phase shifts at the two-photon level [1, 2]. Resch et al. have recently demonstrated somewhat similar effects in a beam-splitter experiment [3]. Because there has been some renewed discussion of this topic, we felt that it would be appropriate to briefly summarize the situation regarding photon exchange interactions. In particular, we note that photon exchange interactions are not useful for quantum information processing because the nonlinear phase shifts that they produce are dependent on the presence of significant photon loss in the form of absorption or scattering. This dependence on photon loss is somewhat analogous to the postselection process inherent in the linear optics approach to quantum computing suggested by Knill, Laflamme, and Milburn (KLM) [4].

Our interest in the use of photon exchange interactions was motivated in part by the fact that there can be a large coupling between a pair of incident photons and the collective modes of a medium containing a large number $N$ of atoms. Under the appropriate phase-matching conditions, this interaction can scale as $N^2$. In addition, the matrix elements for the absorption of two photons and the creation of two excitations of the same collective mode can involve factors of $\sqrt{2}$ that might be expected to give rise to nonlinear effects. Similar nonlinear effects have now been observed by Resch et al in the absorption of photon pairs at an interference filter used as a beam splitter [3].

In our earlier work, we analyzed the effects of photon exchange interactions by assuming that one or more laser beams are used to control the coupling of incident photons with the collective modes of an atomic vapor. The collective modes of the medium can be adequately described by Dicke-state operators $\hat{R}_\pm(\vec{p})$. Here $\hat{R}_+(\vec{p})$ creates a coherent superposition of excited atomic states with a phase factor of $\exp(i\vec{p}\cdot\vec{x}_j)$, where $\vec{p}$ is the wave vector associated with the Dicke state and $\vec{x}_j$ is the position of atom $j$. The operator $\hat{R}_-(\vec{p})$ annihilates the corresponding collective excitation. A laser beam with the appropriate wavelength can be used to control the coupling of an incident photon with wave vector $\vec{k}$ into a Dicke state with wave vector $\vec{p}$ as illustrated by the Raman process of Fig. 1. For photons that are far from resonance from atomic level 3, the effective interaction Hamiltonian has the form

$$H' = \sqrt{N}[\varepsilon_1(t)(\hat{R}_+\hat{a}_1 + \hat{R}_-\hat{a}_1^\dagger) + \varepsilon_2(t)(\hat{R}_+\hat{a}_2 + \hat{R}_-\hat{a}_2^\dagger)] \qquad (1.1)$$

Here $\varepsilon_1(t)$ and $\varepsilon_2(t)$ are time-dependent functions related to the matrix elements and the laser beam intensities while $\hat{a}_1^\dagger$ and $\hat{a}_2^\dagger$ create photons with wave vectors $\vec{k}_1$ and $\vec{k}_2$.

It was assumed that a sequence of laser pulses was applied as illustrated in Fig. 2. Here the system starts out in an initial state $|\phi_0\rangle$ and the laser pulses are used to couple the system through a sequence of other states $|\phi_i\rangle$, possibly including superposition states. The goal was to return the system to the initial state $|\phi_0\rangle$ while maximizing any nonlinear phase shift that was obtained in the process. The probability that the system was left in a final state $|\phi_i\rangle$ will be denoted $P_i$.

We considered a number of laser sequences of that kind, most of which have not been published. In some cases, a more general system was considered in which ancilla photons were incident on the medium as well, or in which additional atomic levels were also included. In all cases, the nonlinear phase shift is zero unless $P_0 \neq 1$, where $P_0$ is the probability of returning to the initial state. The nonlinear phase shift was typically proportional to the probability $P_{loss} = 1 - P_0$ that the system was left in some other final state, which corresponds to photon loss due to absorption or scattering. An example of such a pulse sequence is given in Section V of Ref. 2.

Ref. 2 incorrectly stated that a sequence of pulses could be found in which there was a nonlinear phase shift with no photon loss. The nature of that error has been discussed by Opatrny and Kurizki [5, 6]. Fleischhauer [7] has subsequently given a proof that the nonlinear phase shift must be zero if the system is assumed to return to its initial state.

In the recent experiment by Resch et al [3], two indistinguishable photons were incident on an absorptive beam splitter. A nonlinear increase in the absorption probability was observed when the two photons were incident at the same time. Those results can be interpreted in several ways, including photon exchange interactions. There the role of photon loss is apparent.

KLM [4] have subsequently shown that probabilistic quantum logic operations, including nonlinear phase shifts, can be implemented using linear optical elements and postselection. In retrospect, the need for loss in the photon exchange interaction approach can be seen to be closely related to the use of postselection in the KLM approach. If a pair of photons is incident upon an atomic medium, we will only detect those events in which both photons emerge in the initial state. Other events in which the atoms are left in an excited state, for example, will be rejected simply because the photons will not be detected in that case. If $P_0 \neq 1$, this corresponds to an automatic postselection process in which a nonlinear phase shift could be observed in the remaining two-photon events.

Although the required losses in photon exchange interactions are analogous to the KLM approach in that respect, there are many important differences between the two approaches as well. The use of beam splitters to implement a linear coupling between optical modes in the KLM approach gives rise to a Hamiltonian similar in some respects to Eq. 1 but in a way that is much easier to implement experimentally. In the KLM approach, the probability of a failure event (one that is rejected in the postselection process) can be made arbitrarily small by using a large number of ancilla photons and feed-forward control. And finally, the failure events in the KLM approach can be corrected using a two-qubit encoding, which is essential for the approach to be scalable.

As a result, the KLM linear optics approach is very promising for use in quantum information processing, whereas the photon exchange interaction approach is not.

In summary, there has been some renewed discussion of photon exchange interactions in view of the recent experiment by Resch et al [3]. Our analysis and the more general proof by Fleischhauer [7] both show that photon exchange interactions cannot produce a nonlinear phase shift in the absence of substantial photon loss. Although the need for photon loss is somewhat analogous to the postselection of the KLM approach [4], only the latter is useful for quantum information processing.

This work was funded by ARDA, ARO, and the NSA.

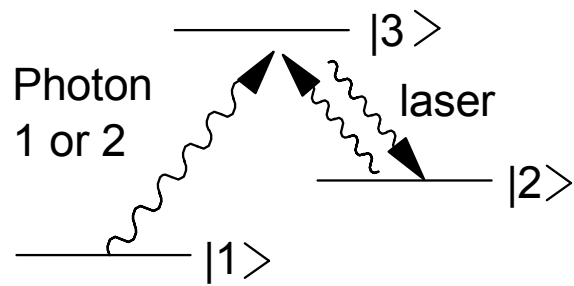

Fig. 1. Raman transition in which a laser beam can be used to control the absorption of a photon and the creation of a collective mode (Dicke state) in an atomic medium.

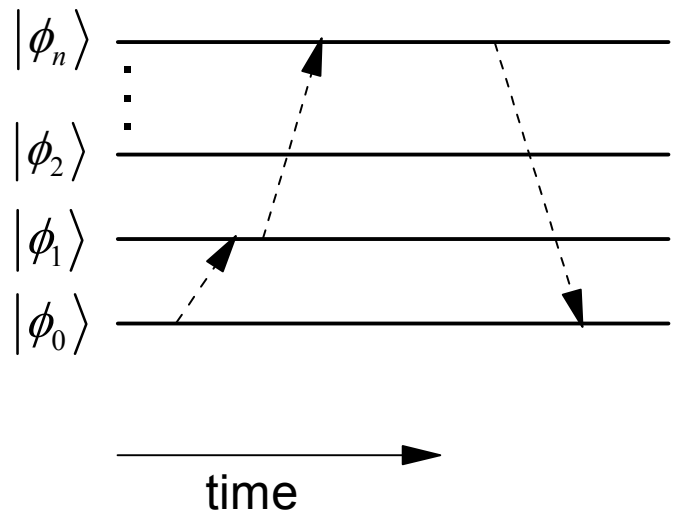

Fig. 2. Use of laser pulses and Raman transitions to control the state of a system composed of two or more incident photons and the collective modes (Dicke states) of an atomic medium. The system is initially in state $|\varphi_0\rangle$ and the goal is to return the system to that state at the end of the process. Only three transitions are shown, but larger numbers of transitions and superposition states may also be used. Nonlinear phase shifts can only be obtained from photon exchange interactions if there is some probability that the system is not in $|\varphi_0\rangle$ at the end of the process.